\documentstyle[preprint,prb,aps]{revtex} 

\newcommand{\sgn}{\mathop{\rm sgn}\nolimits}
\begin{document} 
\draft
\title{ Hall Effect in a Quasi-One-Dimensional System.}
\author{ A.V. Lopatin}
\address{Department of Physics, Rutgers University, Piscataway, NJ 08855}
\date{\today}
\maketitle
\begin{abstract}

 We consider  the Hall effect  in a system of  weakly coupled
one-dimensional chains with
Luttinger interaction within each chain. We construct a perturbation theory 
in the inter-chain
hopping term and find that there is a power law dependence of the
Hall conductivity on the magnetic field with an exponent depending
on the interaction constant. We show that this perturbation theory becomes
valid if the magnetic field is sufficiently large.
 
\end{abstract}
\narrowtext

\section{Introduction}

  It is well known that one-dimensional interacting electrons form a 
non-Fermi-liquid 
system. The Green function acquires an anomalous scaling power $2\Delta$
which implies the breakdown of the basic assumption of the Fermy-liquid theory.
 However, not all correlation functions have anomalous scaling: the correlators
containing only density and current operators are similar to
 those of a noninteracting system. Therefore most of
physical quantities are usual and it is interesting to consider those which
do have a nontrivial contribution from the interaction.

 In this paper we consider the influence of the electron interaction on the
Hall effect in a quasi-one-dimensional system.
To understand the effect of two-dimensionality let us  consider a system of
weakly coupled one-dimensional chains as an example of a quasi-one-dimensional
system. The inter-chain hopping term $t_\perp$ can be neglected for
  electrons with sufficiently
large energy. Therefore the system behaves as one-dimensional on  scales
much smaller than
$l\sim{{\epsilon_F}\over{t_\perp}}a_x$, where $a_x$ is the distance between 
the atoms
within the chains and $\epsilon_F$ is the Fermy energy. 
Actually, the anomalous scaling of Green functions leads to a correction of 
the above 
qualitative expression to 
 $l\sim\left({{t_\perp}\over{\epsilon_F}}\right)^{{-1}\over{1-2\Delta}}a_x$
(see \cite{schulz,braz,bourb}).
On  scales larger than $l$ the system is essentially two-dimensional
and one-dimensional Green functions do not give even qualitatively right 
answer.
Therefore if a physical effect comes from the length scales 
smaller than $l$, then
the answer can   still  have one-dimensional anomalies. We show that the
Hall effect in this system is strongly affected by the
 anomalous powers and acquires a power law
dependence on the magnetic field
$$
\sigma_\perp\sim H^{-1+4\Delta}. 
$$
 This result is valid only if the magnetic field is strong enough. Indeed,
the Hall effect is due to  interference on  scales $l_H\sim
{{\Phi_0}\over{a_yH}}$, where $\Phi_0$ is a quantum of magnetic flux and
$a_y$ is the interchain distance.
We want the system to be  "one dimensional" on these scales, i.e. we need
$l\gg l_H$. Substituting the expression for $l$ we get
\begin{equation}
{{H a_x a_y}\over \Phi_0}\gg \left({{t_\perp}\over{\epsilon_F}}\right)
^{1\over{1-2\Delta}}
\label{est}.
\end{equation}

 To find the Hall conductivity we use  perturbation theory in $t_\perp$.
This perturbation theory becomes valid when the condition (\ref{est}) 
is satisfied
because in this case the effect of hopping term is small on the scales 
 important for  the Hall effect.

 The main
technical difficulty of this problem will be that the Hall conductivity 
of the chains with  the linear  electron spectrum
$$
\epsilon_\pm=\pm v_F(p \mp p_F)
$$
is zero due to the particle-hole symmetry. Therefore it is necessary to 
consider
a nonlinear correction to the spectrum
\begin{equation}  
\epsilon_\pm=\pm v_F(p\mp p_F)+\alpha (p\mp p_F)^2.\label{alphasp}
\end{equation}
We consider the simplest model of spinless electrons. It is also  assumed  that
the system is not close to   half filling so that the  umklapp processes
can be neglected. 

 So  the Hamiltonian of the problem is
\begin{eqnarray}
H=v_{F}\int dx \sum_{i} \hat\psi_i^\dagger\tau_3 
(-i\partial_x)\hat\psi_i-\alpha\int
dx\sum_{i}\hat\psi^\dagger_i\partial_x^2\hat\psi_i \nonumber \\ 
+g\int
dx\sum_i\hat\psi_{i+}^\dagger\hat\psi_{i+}\hat\psi_{i-}^\dagger\hat\psi_{i-}
+t_\perp\int dx \sum_{<i,j>} \hat\psi^\dagger_i\hat\psi_j 
e^{-i{e\over c}A_{i,j}}
\label{ham0},
\end{eqnarray}
where $\hat\psi$ is composed from the right- and left-moving electrons
$\hat\psi=\left(\begin{array}{c} \hat\psi_+ 
\\ \hat \psi_- \end{array}\right) $ 
, $\tau_3$ is a Pauli matrix, 
 $A_{i,j}=\int_i^{j}{\bf A} d{\bf l}$ and we  use the
Landau gauge $A_y=Hx$.
The second term in the Hamiltonian is the nonlinear correction to the free
electron spectrum.
The model without  the hopping term and $\alpha-$term  
can be solved exactly, for example
by the bosonization method. It is also possible  to bosonize the 
 $\alpha$-term \cite{haldane} but it leads to a
  cubic interaction between the bosons and the model with
such an interaction is not exactly solvable. Therefore we have to consider 
the $\alpha-$term  as a perturbation, too. The nonlinear term
in the spectrum (\ref{alphasp}) 
is small compared with the linear term 
 on  scales bigger than ${{\alpha}\over {v_F}}\sim a$ ,
hence a perturbation
theory in $\alpha$ is valid if $l_H\gg a.$  This condition can be written as
$$
{{Ha_x a_y}\over{\Phi_0}}\ll 1,
$$
and it is satisfied for any real experimental situation. Therefore
a perturbation theory in $\alpha$ is always a good approximation.

The plan of the paper is the following: In Section \ref{two} we express
 the Hall conductivity through the single-chain correlation functions.
In Section \ref{three} we explain the technique that we will use
to calculate  one-dimensional
correlators. In Section \ref{four} we calculate the Hall conductivity.
Finally, we summarize our results and discuss their
possible applications in Section\ref{five}.

   \section{ Expression for the  Hall conductivity .}\label{two}

   Let us choose a coordinate frame so that the magnetic field points
 along the z-axis and the chains are along the x-axis. The label $i$, which
 denotes the number of the chain,  increases in the y-direction.
 The electric field $E_x=E_0 e^{-i\omega t}$ is applied along the 
 chains (x-axis). In this geometry the Hall conductivity $\sigma_\perp(\omega)$
relates  the electric field $E_x$ with the current between the chains $j_y$
 $$j_y=\sigma_\perp(\omega) E_x .$$ 
 According to the Kubo formula the conductivity is expressed through
 the retarded current-current correlator
\begin{equation}
\sigma_\perp(\omega)={1\over\omega}\sum_i\int dx P_R(x,i,\omega),
\end{equation}
where
\begin{equation}
P_R(x,i,\omega)=
\int_{-\infty}^{\infty}dt\, e^{i\omega t}
[\hat j_y (x,i,t),\hat j_x(0,0,0)]\theta(t).
\label{ret}
\end{equation}
 In these expressions $\hat j_y$ is the  Heisenberg operator of the
 current between the chains
\begin{equation}
\hat j_y(x,i,0)=t_\perp ei\left(\hat\psi^\dagger_i(x)\hat\psi_{i+1}(x) 
e^{-i{e\over c}
A_{i,i+1}}-h.c.\right),
\end{equation}
and $\hat j_x$ is the in-chain current operator
\begin{equation}
\hat
j_x(x,i,0)=e\left(v_F\hat\psi_i^\dagger(x)\tau_3\hat\psi_i(x)
+2\alpha\hat\psi_i^\dagger(x)(-i\partial_x)
\hat\psi_i(x)\right).
\end{equation}
We will work in the Euclidean space which corresponds to 
the  Wick rotation $t\to -it$,
$\omega\to i\omega$  (See for example \cite{peskin}). 
The Euclidean Lagrangian of this problem is
\begin{eqnarray}
L_{\cal E}=
\int dx\int dt \sum_{i} \psi_i^\dagger(-\partial_0+iv_F\tau_3 \partial_1+\alpha
\partial_1^2)\psi_i
-g\int dx\int dt\sum_i
\psi_{i+}^\dagger\psi_{i+}\psi_{i-}^\dagger\psi_{i-} \nonumber\\
-t_\perp\int dx \int dt\sum_{<i,j>} \psi^\dagger_i\psi_j 
e^{-i{e\over c}A_{i,j}}
 , \label{lag}
\end{eqnarray} 
where   $\psi^\dagger,\psi$ are two-vector
 anticommuting variables which correspond to the operators 
$\hat\psi^\dagger,\hat\psi$ 
of the Hamiltonian formalism and $\partial_0=
{\partial\over{\partial t}},\partial_1={\partial\over{\partial x}}$. 
Let us define the Euclidean current-current correlator
\begin{equation}
P_{\cal E}(x,i,\omega)=
\int_{-\infty}^{\infty}dt\, e^{i\omega t}\langle T_t \hat j_y
(x,i,t),\hat j_x(0,0,0)\rangle
\end{equation}
where $T_t$ means time ordering in the "Euclidean" time. For 
Euclidean correlators
the standard perturbation theory can be used. The retarded correlator 
(\ref{ret})
is the analytical continuation of the Euclidean one
\begin{equation}
P_R(\omega)=-iP_{\cal E}(-i\omega).
\end{equation}

Considering the last term in (\ref{lag}) as a  perturbation, 
 one can find the Euclidean current-current correlator
to the first order in $t_\perp$
\begin{eqnarray}
\int dx_3\sum_{i_3}\langle j_x(x_3,i_3,t_3) j_y(x_1,i_1,t_1)
\rangle_{L_{\cal E}}
\nonumber \\
=2et_\perp^2\sum_{s}\int dx_3\int dx_2\int dt_2 
\Bigl[\langle j(x_3,t_3)\psi_s(x_1,t_1)\psi^\dagger_s(x_2,t_2)\rangle
\langle \psi_s(x_2,t_2)\psi_s^\dagger(x_1,t_1)\rangle \nonumber \\
+(x_1,t_1\leftrightarrow x_2,t_2)\Bigr]
\sin q(x_1-x_2) \label{corr}         \end{eqnarray}
where $q={{eHa_y}\over c}$ and
the label $s=+,-$.
Note that the one-dimensional current $j$ contains a contribution 
from the nonlinear 
correction to the
spectrum
\begin{equation}
j=j^{(0)}+j^{(1)}, \label{onecur}
\end{equation}
where
\begin{eqnarray}
j^{(0)}=ev_F\psi^\dagger\tau_3\psi \\
j^{(1)}=2e\alpha \psi^\dagger(-i\partial_1)\psi.
\end{eqnarray}
 The angular brackets on the r.h.s. of (\ref{corr}) represent averaging with
respect to
  the single-chain Lagrangian
\begin{equation}
L_{\cal E}^{(1)}=
\int dx\int dt\left(\psi^\dagger(-\partial_0+iv_F\tau_3 \partial_1+\alpha
\partial_1^2)\psi
-g\psi_+^\dagger\psi_+\psi^\dagger_-\psi_-\right) . \label{l1}
\end{equation}
Finally for the Hall conductivity we have the following expression
\begin{equation}
\sigma_\perp(\omega)=-{{2eit_\perp^2 }\over \omega} \Gamma_{\cal E}(-i\omega)
\label{sigma},
\end{equation}
where $\Gamma_{\cal E}(\omega)$
 is the Euclidean correlator
$$\Gamma_{\cal E}(\omega)=\int _{-\infty}^\infty dte^{i\omega t}
\Gamma_{\cal E}(t), $$
with
\begin{eqnarray}
\Gamma_{\cal E}(t_1-t_3)  \nonumber \\
=\sum_{s}\int dx_3\int dx_2 \int dt_2
\Bigl[\langle j(x_3,t_3)\psi_s(x_1,t_1)\psi^\dagger_s(x_2,t_2)\rangle 
\langle \psi_s(x_2,t_2)\psi_s^\dagger(x_1,t_1)\rangle \nonumber \\
+
(x_1,t_1\leftrightarrow x_2,t_2)\Bigr]
\sin q(x_1-x_2).  \label{gamma}
\end{eqnarray}

\section{Bosonization.}\label{three}

Now the problem is to find the one-dimensional correlators in (\ref{gamma}).
The standard way to treat a one-dimensional model
with the linear spectrum  is bosonization.
 Without the 
 $\alpha$-term the model (\ref{l1}) can be solved exactly.
We will treat this term as a perturbation and begin with bosonization
 of  the Hamiltonian. The $\alpha$-term in
the fermionic Lagrangian
results in a cubic interaction between the bosons \cite{haldane}.
 The single-chain  Hamiltonian
$ H_1 $ in the bosonized form 
  (see Appendix A) is
\begin{equation}
\hat H_1={1\over 2}\left((\partial_1\hat \Phi)^2
+\hat\Pi^2\right)
+{\alpha\over 3}\beta\partial_1\hat\Phi
\left(3{\pi\over{\beta^2}}\hat\Pi^2+{{\beta^2}\over\pi}
(\partial_1\hat\Phi)^2\right),
\label{ham}
\end{equation}
where $\Phi$ and $\Pi$ are canonically conjugate Bose-operators and $\beta$ is
a constant which depends on the interaction constant $g$
$$
\beta=\sqrt{\pi}\left({{v_F-g/2\pi}\over{v_F+g/2\pi}}\right)^{1\over 4}.
$$
 The fermionic  and current operators  (see Appendix A) are
\begin{eqnarray}
\hat\psi_{\pm}(x,t)={1\over{\sqrt{2\pi\delta}}}
e^{\pm i\hat\Phi_\pm(x,t)},\,\,\,\,
\hat\Phi_\pm(x,t)=\beta\hat\Phi(x,t)
\mp{\pi\over\beta}\int_{-\infty}^x dx^\prime\hat\Pi(x^\prime,t) \label{psi}\\
\hat j=-e({{v_F}\over\beta}\hat\Pi+2\alpha  \hat \Pi\partial_1\hat\Phi )
\label{carrent},
\end{eqnarray}
where $\delta$ is the inverse momentum-space cutoff, 
and the normal ordering of operators is implied in (\ref{ham}) 
and (\ref{carrent}). 
 Here we have rescaled the energy units so that the velocity of 
the Bose-particles $v_B$
 is $1$.
The second term in the current operator is due to the $\alpha$-term
in the Hamiltonian.
 It is more convenient to calculate the correlation functions using
the functional representation. 
 In this formalism we introduce the fields $\Phi,\Pi$ which are
related to the fermion fields $\psi^\dagger,\psi$ via
\begin{eqnarray}
\psi_{\pm}(x,t)={1\over{\sqrt{2\pi\delta}}}e^{\pm i\Phi_\pm(x,t)}, \,\,\,\,
\Phi_\pm(x,t)=\beta\Phi(x,t)
\mp{\pi\over\beta}\int_{-\infty}^x dx^\prime\Pi(x^\prime,t) \\
 j=-e({{v_F}\over\beta}\Pi+2\alpha   \Pi\partial_1\Phi  ).
\end{eqnarray}
The  Green
function  
\begin{equation}
G(x_1-x_2,t_1-t_2)=\langle T_t\hat\psi(x_1,t_1)
\hat\psi^\dagger(x_2,t_2)\rangle_H,
\\ \label{Gh}
\end{equation}
calculated by the  Hamiltonian method is related to
 the Green function 
$$
G_f(x_1-x_2,t_1-t_2)=\langle \psi(x_1,t_1)\psi^*(x_2,t_2)\rangle_f,
$$
calculated by the functional method through the following relation
\begin{equation}
G(x_1-x_2,t_1-t_2)=\sgn(t_1-t_2)G_f(x_1-x_2,t_1-t_2). \label{sgn}
\end{equation}
 Indeed, due to the 
fundamental property of the correspondence between the Hamiltonian 
and functional methods
we have
$$
\langle T^{(boz)}_t\hat\psi(x_1,t_1)\hat\psi^\dagger(x_2,t_2)\rangle_H=
\langle \psi(x_1,t_1)\psi^*(x_2,t_2)\rangle_f,
$$
where $T_t^{(boz)}$ means time-ordering of the boson operators because the
 $\hat\psi$-operators
are constructed from the boson operators.
But  the definition (\ref{Gh}) implies 
 time-ordering of  fermions, and  this difference should be corrected by
 inserting
$\sgn(t_1-t_2)$ into (\ref{sgn}).
So to get an actual "Hamiltonian" Green function we should multiply
the corresponding "functional" Green function by $\sgn(t_1-t_2).$

The  Lagrangian density corresponding to the single-chain Hamiltonian 
(\ref{ham})
 in the Euclidean space is
\begin{equation}
{\cal L}^{(1)}_{\cal E}={\cal L}^{(0)}_{\cal E}+V(\Pi,\partial_1\Phi) 
\label{lone},
\end{equation}
where 
\begin{eqnarray}
{\cal L}^{(0)}_{\cal E}=
i\Pi\partial_0\Phi-{1\over 2}\left(\Pi^2+(\partial_1\Phi)^2\right),\\
V(\Pi,\partial_1\Phi)=-{\alpha\over
3}\beta\partial_1\Phi\left(3{\pi\over{\beta^2}}\Pi^2+
{{\beta^2}\over\pi}(\partial_1\phi)^2\right).
\end{eqnarray}
We will treat the interaction $V$ as a perturbation. The bare correlation 
functions are
 \begin{eqnarray}
\langle\Phi(x_1,t_1)\Phi(x_2,t_2)\rangle^{(0)}_f=
-d(x,t) \label{cor1}\\
\int_{-\infty}^{x_1}dx^\prime\int_{-\infty}^{x_2}dx^{\prime\prime}
\langle\Pi(x^\prime,t_1)\Pi(x^{\prime\prime},t_2)\rangle^{(0)}_f=
-d(x,t)
 \label{cor2}\\
\int_{-\infty}^{x_1}dx^\prime\langle\Pi(x^\prime,t_1)\Phi(x_2,t_2)
\rangle^{(0)}_f=
-{i\over{2\pi}}\sgn t\left(\arctan{x\over{|t|+\delta}}+{\pi\over 2}\right), 
\label{cor3}
\end{eqnarray}
where
\begin{equation}
d(x,t)={1\over {4\pi}}\ln{{(|t|+\delta)^2+x^2}\over{ l^2}}. \label{greenf}
\end{equation}
In the above formulas $t=t_1-t_2$ and $x=x_1-x_2$. The label $(0)$ in
 (\ref{cor1}-\ref{cor3})
means averaging with respect to the  free-boson Lagrangian 
$L^{(0)}_{\cal E}$ and $l$ is
the length of the chains.
Using the correlators (\ref{cor1}-\ref{cor3}) we can find the 
 correlators of $\Phi_{\pm}$:
\begin{eqnarray}
\langle \Phi_{\pm}(x_1,t_1)\Phi_{\pm}(x_2,t_2)\rangle^{(0)}-
\langle \Phi_{\pm}(0,0)\Phi_{\pm}(0,0)\rangle^{(0)} \\ \nonumber
=-{1\over 4}\left({\pi\over {\beta^2}}+{{\beta^2}\over \pi}\right)
\ln{{t^2+x^2}\over{\delta^2}}\pm i\arctan{x\over t}.
\end{eqnarray}
Here and below we put $\delta=0$ wherever it does not lead to apparent 
divergences.
  
Now let us reproduce the well known result for the one-particle Green function.
 According to the formula
\begin{equation}
\langle e^A\rangle=e^{{1\over 2}\langle A^2\rangle}, \label{ea}
\end{equation}
where $A$ can be any linear functional of the fields $\Pi, \Phi$, we can 
easily find
the correlators of  $\psi$-functionals
\begin{equation}
G_{\pm,f}^{(0)}(\xi_1-\xi_2)=
\langle \psi_{\pm}(\xi_1)\psi^*_{\pm}(\xi_2)\rangle^{(0)}_f=
\sgn(t_1-t_2){i\over{2\pi}}{1\over{\pm x+i
t}}\left({{\delta^2}\over{x^2+t^2}}\right)^\Delta, \label{gf0}
\end{equation}
where $\Delta={1\over 4}\left({\pi\over {\beta^2}}+{{\beta^2}\over \pi}-
2\right)$
and $\xi=(x,t)$.
The appearance of the factor $\sgn(t_1-t_2)$ is in  accordance 
with the formula (\ref{sgn}).
Indeed, due to this formula the fermion Green function with proper 
time-ordering is
\begin{equation}
G_{\pm}^{(0)}(\xi_1-\xi_2)=
\left\langle \hat
T_t\hat\psi_{\pm}(x_1,t_1)\hat\psi^{\dagger}_{\pm}(x_2,t_2)
\right\rangle^{(0)}_H=
{i\over{2\pi}}{1\over{\pm x+i t}}\left({{\delta^2}\over{x^2+t^2}}\right)^\Delta
\label{gh0}
\end{equation}
which is the right answer.

 It will be convenient for us to introduce the following generating functional
\begin{equation}
Z_\pm(f_0,f_1)=\langle \psi_\pm(\xi_1)\psi_\pm^*(\xi_2) e^{
\int d^2\xi\left[f_0(\xi)\Pi(\xi)+f_1(\xi)\partial_1\Phi(\xi)\right]}
\rangle^{(0)}_f.
\label{z0}
\end{equation}
 This functional can be  calculated using the formula (\ref{ea})
\begin{equation}
Z_{\pm}(f_0,f_1)=G^{(0)}_{\pm,f}(\xi_1,\xi_2)e^{F^{\pm}(f_0,f_1)+{1\over 2}
\int d^2 \xi_1d^2\xi_2
f^T(\xi_1)D(\xi_1,\xi_2)f(\xi_2)}. \label{zf}
\end{equation}
In this formula
 $G^{(0)}_{\pm,f}$ is the Green function (\ref{gf0}),
$f=(f_0,f_1)^T$ is a two-vector constructed from $f_0,f_1$, and
 $F^{\pm}(f_0,f_1)$ is a linear functional of $f_0,f_1$
\begin{equation}
F^{\pm}(f_0,f_1)= \int d^2\xi
\Bigl(f_0(\xi)J_0^{\pm}(\xi_1,\xi_2,\xi)+f_1(\xi)J_1^{\pm}(\xi_1,\xi_2,\xi)
\Bigr) , 
\end{equation}
where
\begin{eqnarray}
J_0^{\pm}(\xi_1,\xi_2,\xi_3)={1\over {2\pi}}
{{-i{\pi\over\beta}(x_1-x_3)\mp\beta (t_1-t_3)}\over
{(t_1-t_3)^2+(x_1-x_3)^2}} -(x_1,t_1\leftrightarrow x_2,t_2),\label{j1} \\ 
J_1^{\pm}(\xi_1,\xi_2,\xi_3)={1\over {2\pi}}
{{\pm i\beta(x_1-x_3)+{\pi\over\beta}(t_1-t_3)}\over
{(t_1-t_3)^2+(x_1-x_3)^2}}-(x_1,t_1\leftrightarrow x_2,t_2) \label{j2}.
\end{eqnarray}
 Finally, $D$
is a matrix  Green function which in the momentum
space is
\begin{equation}
D(\omega,p)={{p^2}\over{p^2+\omega^2}}\left[ \begin{array}{cc}
1&i{\omega\over p}\\i{\omega\over p}&1 
\end{array} \right]. \label{gd}
\end{equation}

So far we have calculated the generating functional which corresponds to
the  free-boson Lagrangian $L^{(0)}_{\cal E}$.
 But the  averaging in (\ref{gamma})
 is done with respect to 
 the Lagrangian $L^{(1)}_{\cal E}$ (\ref{lone}),
and therefore  we need the following
generating functional
$$
Z^{-1}\int D\Pi D\Phi\,\,\psi_{\pm}(\xi_1)\psi^*_{\pm}(\xi_2)
e^{ L_{\cal E}^{(1)}+
\int d^2\xi\left[f_0(\xi)\Pi(\xi)+f_1(\xi)\partial_1\Phi(\xi)\right]},
$$
where
$$
Z=\int D\Pi D\Phi\,e^{ L_{\cal E}^{(1)}}.
$$
 Before doing the perturbation theory in $V$ it is convenient to
 shift the fields $\Pi\to\Pi+J_0^{\pm},\,\,\partial_1\Phi\to\partial_1
\Phi+J_1^{\pm}$
 and  redefine them $\Pi=\gamma_0,\partial_1\Phi=\gamma_1$:
\begin{eqnarray}
Z^{-1}\int D\Pi\, D\Phi\,\,\psi_{\pm}(\xi_1)\psi^*_{\pm}(\xi_2)
e^{L_{\cal E}^{(1)}+
\int d^2\xi (f_0\Pi+f_1\partial_1\Phi)}  \nonumber \\
=G^{(0)}_{\pm,f}(\xi_1-\xi_2)
Z_\gamma^{-1}
\int D\gamma_0 D\gamma_1\, e^{\int d^2\xi [-{1\over 2}\gamma^T
\hat D^{-1}\gamma
+V(\gamma_0+J_0,\gamma_1+J_1)+f_0(\gamma_0+J_0)
+f_1(\gamma_1+J_1)]}. \label{z}
\end{eqnarray}
In this representation $V$ is the interaction term 
$$
V(\gamma_0,\gamma_1)=
-{\alpha\over 3}\beta\,\gamma_1\left({3\pi\over{\beta^2}}\gamma_0^2+
{{\beta^2}\over\pi}\gamma_1^2\right),
$$
and 
$$
Z_\gamma=\int D\gamma_0 D\gamma_1\, e^{\int d^2\xi [-{1\over 2}\gamma^T
\hat D^{-1}\gamma
+V(\gamma_0,\gamma_1)]}.
$$
The arguments of the $J$-functions in (\ref{z}) are $J(\xi_1,\xi_2,\xi).$
The field $\gamma$ is a two-component field $\gamma=(\gamma_0,\gamma_1)^T$
and $\hat D^{-1}$ is an integral
operator corresponding to the Green function $D^{-1}$, which in the 
momentum space is the inverse Green function (\ref{gd})
$$
D^{-1}(\omega,p)=\left[ \begin{array}{cc}
1&-i{\omega\over p}\\-i{\omega\over p}&1 
\end{array} \right].
$$
 This representation is convenient for the perturbation theory in $\alpha$
because it already contains the Green function $G_\pm^{(0)}$ so that we 
do not have to  deal with $\psi$-functionals any more.
 To prove the representation (\ref{z}) it is convenient to use the
 following formulas for
$J-$functions:

\begin{eqnarray}
J_0^{\pm}(\xi_1,\xi_2,\xi)=\left[i{\pi\over \beta}{\partial\over{\partial x}}
\pm \beta{\partial\over{\partial t}}\right]
\left[d(\xi_1-\xi)-d(\xi_2-\xi)\right], \\
J_1^{\pm}(\xi_1,\xi_2,\xi)=\left[\mp \beta i{\partial\over{\partial x}}
-{\pi\over{\beta}}{\partial\over{\partial t}}\right]
\left[d(\xi_1-\xi)-d(\xi_2-\xi)\right], \\
 {[\partial_0^2+\partial_1^2]}d(\xi)=\delta(\xi)
\end{eqnarray}

where $ d(\xi)$ is defined in (\ref{greenf}).

\section{ Calculation of the Hall conductivity.}

\label{four}

To calculate $\Gamma$ we need to find the correlators
$$K_\pm(\xi_1,\xi_2,\xi_3)=\int dx_3\langle j (\xi_3)\psi_\pm(\xi_1)
\psi_\pm^\dagger(\xi_2)\rangle $$ and
$$G_\pm(\xi)=\langle \psi(\xi) \psi^\dagger(0)\rangle,$$
where the  averaging is done with respect the single-chain Lagrangian 
(\ref{l1}). 
Let us apply the technique developed in the 
previous section to calculate these correlators.
First of all we should find these correlators
to the zeroth order in $\alpha$.
The Green function $G_{\pm}^{(0)}$ was already calculated in the 
previous section. The correlator $K_{\pm}$ to the zeroth order in $\alpha$
can be calculated using the generating functional (\ref{z0},\ref{zf})
$$
K_{\pm}^{(0)}=-{{e v_F}\over \beta}\sgn(t_1-t_2)\int dx_3
{\delta\over{\delta f_0}}Z_\pm(f_0,f_1)_{|f_0=f_1=0}$$ 
$$=
-{{e v_F}\over\beta}
\int dx_3 G_{\pm}^{(0)}(\xi_1,\xi_2) J_0(\xi_1,\xi_2,\xi_3).
$$
Calculating the integral over $x_3$ we get
\begin{eqnarray}
K_\pm^{(0)}=\int dx_3\langle j_0(\xi_3)\psi_\pm(\xi_1)
\psi^\dagger_\pm(\xi_2)\rangle^{(0)}  \nonumber\\
=\mp{1\over 2}e v_F G_\pm^{(0)}(\xi_1-\xi_2)
\bigl[\sgn(t_3-t_1)-\sgn(t_3-t_2)\bigr].
\end{eqnarray}
Substituting these expressions and the Green function (\ref{gh0})
 into $\Gamma$ (\ref{gamma}) we get zero due to the fact 
that $G_\pm^{(0)}(x_1-x_2,t_1-t_2)$ is odd under the transformation
$x_1,t_1\leftrightarrow x_2,t_2$ and $K^{(0)}_\pm$ is even. Indeed, the model
 with
the linear spectrum 
($\alpha=0$) should give  zero  Hall effect due to the particle-hole symmetry. 
 To the first order in $\alpha$ we can conveniently represent the product 
of these two correlators as
\begin{equation}
\langle j\psi\psi^\dagger\rangle\langle\psi\psi^\dagger\rangle=
\langle j_0\psi\psi^\dagger\rangle\langle\psi\psi^\dagger\rangle+
\langle j_1\psi\psi^\dagger\rangle^{(0)}\langle
\psi\psi^\dagger\rangle^{(0)},\label{pre}
\end{equation}
where $j_0$ and $j_1$ are defined in (\ref{onecur}).
The first term in (\ref{pre}) contains the current to the
 zeroth order so that the nonlinear corrections
come from the Lagrangian. The second term contains no $\alpha$-corrections from
the Lagrangian because $j_1$ is already proportional to $\alpha$. 
Fortunately the first term gives no contribution when substituted into the 
formula
for $\Gamma$ (see Appendix B) and therefore we should calculate only the 
contribution
from the second term. The  correlator
$$
\int dx_3\langle j_1(\xi_3)\psi_{\pm}(\xi_1)\psi_{\pm}(\xi_2)\rangle^{(0)},
$$
where the current correction $j_1$ in the bosonized form is
$$
j_1=-2e\alpha\Pi\,\partial_1\Phi,
$$
 can be calculated using the formulas (\ref{z0},\ref{zf})
$$
\int dx_3\langle j_1(\xi_3)\psi_{\pm}(\xi_1)\psi_{\pm}(\xi_2)\rangle^{(0)} $$
$$
=-2e\alpha\int dx_3 J_0^\pm(\xi_1,\xi_2,\xi_3)J_1^\pm(\xi_1,\xi_2,\xi_3)
G_{\pm}^{(0)}(\xi_1,\xi_2),
$$
where it was used that  $\int dx_1 D(\xi_1,\xi_2)=0$ ( see Appendix B).
Taking the integral over $x_3$  we get 

\begin{eqnarray}
\int dx_3\langle j_1(\xi_3)\psi_\pm(\xi_1)\psi^\dagger_\pm(\xi_2)\rangle= 
{{\alpha e i}\over{2\pi}} G_{\pm}^{(0)}(\xi_1-\xi_2)
{1\over{(x_1-x_2)^2+(t_1-t_2)^2}} \nonumber \\
\left[
\pm2\pi i(t_1-t_2)-\left((\pi/\beta)^2+\beta^2\right)(x_1-x_2)\right]
\bigl[\sgn(t_3-t_1)-\sgn(t_3-t_2)\bigr]. \label{intk}
\end{eqnarray}
Substituting (\ref{intk}) and (\ref{gh0}) into the expression for $\Gamma$ 
 and 
 shifting the variables $x_2\to x_2+x_1,
t_2\to t_2+t_1$ for $\Gamma_{\cal E}(\omega)$ we get
 
\begin{equation}
\Gamma_{\cal E}(\omega)=
{{e\alpha i}\over{2\pi^3}}\delta^{4\Delta}\int dt^\prime\int dt\, 
e^{i\omega t}\left[
\sgn t-\sgn (t^\prime+t)\right]f(t^\prime),
\end{equation}
where
$$
f(t)=\int dx{1\over{(x+it)^2}}{1\over{(x^2+t^2)^{1+2\Delta}}} 
\left[-2\pi i t+x\left((\pi/\beta)^2+\beta^2\right)\right] \sin q x.
$$
Using  that $f(t^\prime)$ is an even function of $t^\prime$
and integrating by parts one can show that
$$
\int dt^\prime\int dt \,e^{i\omega t}\left[\sgn t-\sgn(t+t^\prime)\right]
f(t^\prime) $$
$$
=-{{4i}\over\omega}\int_0^\infty dt (\cos\omega t -1)f(t).
$$
Therefore 
\begin{equation}
\Gamma_{\cal E}(\omega)=
{{2e\alpha}\over{\omega\pi^3}}\delta^{4\Delta}\int_0^\infty dt 
(\cos\omega t-1)f(t).
\end{equation}
Introducing cylindrical coordinates $x_2=\rho \cos\phi, t=\rho \sin\phi$ 
we can integrate over $\phi$ obtaining
\begin{eqnarray}
\Gamma_{\cal E}(\omega)=
{{e
\alpha}\over{\pi^2\omega}}\delta^{4\Delta}
\int_0^\infty{{d\rho}\over{\rho^{2(1+2\Delta)}}}
\biggl [(\pi/\beta-\beta)^2{q\over{\sqrt{\omega^2+q^2}}}J_1(\rho
\sqrt{\omega^2+q^2})
\nonumber \\
-  (\pi/\beta+\beta)^2{{ q(q^2-3\omega^2)}\over{(\omega^2+q^2)^{3/2} }}
J_3(\rho\sqrt{\omega^2+q^2})-(\omega=0)\biggr]\label{ge},
\end{eqnarray}
where $J_1$ and $J_3$ are the Bessel functions (not to be confused with
$J$-functions (\ref{j1},\ref{j2})).
Calculating the integral over $\rho$ and using (\ref{sigma}) for the
Hall conductivity we get
\begin{equation}
\sigma_\perp(\omega)=-{{2e^2t_\perp^2\alpha}\over\pi}\left({\delta\over 2}
\right)^{4\Delta}
{{\Gamma(1-2\Delta)}\over{\Gamma(2+2\Delta)}}
{q\over{\omega^2}}
\left[(q^2-\omega^2)^{2\Delta}{{q^2+\omega^2}\over{q^2-\omega^2}}-q^{4\Delta}
\right],
\end{equation}
which is the main result of this paper.
 Expanding in $\omega$ we get the DC-conductivity 
\begin{equation}
\sigma_\perp=-{4\over\pi}{{\Gamma(1-2\Delta)}\over{\Gamma(2+2\Delta)}}
(1-\Delta){{t^2_\perp e^2\alpha}\over{v_B^2\left({{eHa_yv_B}\over c}
\right)^{1-4\Delta}}} \left({\delta\over { 2 v_B}}\right)^{4\Delta}, 
\label{ans}
\end{equation}
where $v_B$ was restored.
One can see that $\sigma_\perp$ depends on the magnetic field as 
$H^{4\Delta-1}$.
Note that the formula (\ref{ans}) does not work for $\Delta>{1\over 2}$.
(The integral over $\rho$ in (\ref{ge}) does not converge in this case.)
The fact that $\Delta ={1\over 2}$ is a critical value is not surprising
because it was shown \cite{wen} that in the case  $\Delta>{1\over 2}$ 
the interchain hopping term in the
Hamiltonian is  irrelevant in RG sense. To avoid confusion we note
that the irrelevance of the hopping term in the case $\Delta>{1\over 2}
$ should be
understood in the straightforward dimensional RG sense. Actually, 
considering two-particle tunneling processes, one can see that
the  hopping term is relevant even for $\Delta >{1\over 2}$
(See \cite{braz,bourb,yakov}).
As was mentioned above, this theory is valid for high magnetic fields. To 
find the 
 applicability criterion we should consider the next nonzero  order in 
$t_\perp.$ 
It is
hard to calculate it but  from  dimensional analysis it follows that the 
correction should have the form
$$
\sigma_\perp\to\sigma_\perp\left(1+const\cdot
{{t^2_\perp}\over{\left({{eHa_y v_F}\over c}
\right)^2}}\left({{eHa_y\delta}\over c}\right)^{4\Delta}\right).
$$
Indeed, expanding to the next nonzero order in $t_\perp$  one gets
four additional $\psi-$operators which give the factor $\delta^{4\Delta}$ 
and  the 
remaining factors
can be restored from  dimensionality.
This correction is small if
$$
H\gg{{\Phi_0}\over{a_x a_y}}\left({{t_\perp}\over {\epsilon_F}}\right)^
{1\over{1-2\Delta}},
$$
where $\Phi_0$ is a quantum of  magnetic flux. It agrees with our expectation
(\ref{est}) in the Introduction.

  \section{ Discussion and conclusions.} \label{five}

 We considered the Hall effect in  a quasi-one-dimensional interacting 
electron system.
It was found that in high magnetic fields
\begin{equation}
H\gg{{\Phi_0}\over {a_x a_y}}\left({{t_\perp}\over {\epsilon_F}}
\right)^{1\over{1-2\Delta}}
\label{condit}
\end{equation}
there is a power law dependence of the Hall conductivity on the magnetic field 
\begin{equation}
\sigma_\perp\sim H^{-1+4\Delta} \label{res}
, \end{equation}
where $2\Delta$ is the anomalous exponent of the one-dimensional Green function
(\ref{gh0}).
This formula can be applied for $\Delta<{1\over 2}$. This result was 
obtained for
the zero-temperature case. But for nonzero temperatures  much lower
 than ${{eHa_x v_F}\over c}$
 it should still hold because the temperature will change
the correlation functions only in the low-energy 
region which is irrelevant in the case of high magnetic fields. 
Therefore this  result  can be applied for nonzero  temperatures if
\begin{equation}
 {T\over {\epsilon_F}}\ll {{Ha_x a_y}\over \Phi_0}.
\end{equation}
 We considered the simplest model of spinless electrons. In a more realistic
case of the Hubbard model the single-chain Green function has a more  
complicated form
\begin{equation}
G(x,t)=
{i\over{2\pi}}{1\over{(\pm x+v_\rho it)^{1/2}(\pm x+v_s it)^{1/2} }}
\left({{\delta^2}\over {x^2+(v_\rho t)^2}}\right)^\Delta \label{hab}
,\end{equation}
 (in the Euclidean space),
where $v_s$ and $v_\rho$ are the spin and charge velocities.
Using dimensional analysis one can argue that
the result (\ref{res}) should  still hold. Indeed, the formula (\ref{corr})
 is right in this case. Comparing this formula with the 
final answer,  one can see that every $\psi$-operator gives rise to
 an additional factor $\delta^\Delta$ in the final answer, while the 
current $j$ 
gives no additional powers  because $j$
 is a local function of the Bose-fields. Therefore
the exponent of the magnetic field in the final answer should be the same 
with $\Delta$
defined by (\ref{hab}).

We hope that experimentally this effect can be observed in  one-dimensional 
organic
conductors (see \cite{jerome} for review). Actually, the possibility of 
observing
 this effect  depends
on the value of $\Delta $ in a particular material. If we take 
$a_x a_y=5\times 10^{-15} cm^2 $ and ${{t_\perp}\over{\epsilon_F}}={1\over 10}$
 which is typical for these 
materials, then we can rewrite the condition (\ref{condit}) in the form
$$
H\gg (10^3\, ...\,\,10^4)\left({1\over{10}}\right)^{1\over{1-2\Delta}} \,\,\,T.
$$
 It can  be satisfied from the experimental point of view if $\Delta$ is not
far from ${1\over 2}.$
Note that one-dimensional organic conductors at low temperatures typically
exhibit phase transitions to superconducting, CDW or SDW states and 
 many properties such as Hall effect, magnetoresistance, etc. are very
unusual at these temperatures. Of course our result  can  not be
applied in these cases because effects of complicated ground state were
not considered. In other words, for this effect to be observed
the materials must be in the metallic state.

 \section*{Acknowledgments}

 I am indebted to  L. B. Ioffe for the idea of this work and very 
useful discussions.

 \section*{ Appendix A. Bosonization of the single-chain Hamiltonian.}

 In this section we will follow the approach described in \cite{shankar}.
 In the bosonization technique
the product of two operators which has an infinite vacuum matrix element
$$ \langle \hat A \hat B \rangle \to \infty $$
should be understood as a limit
$$
\langle \hat A(x) \hat B(x) \rangle  \to
\langle \hat A(x) \hat B(x^\prime) \rangle_{x\to x^\prime}.
$$
Usually the r.h.s. of this expression can be written as
$$
{z\over{(x-x^\prime)^\eta}}+\hat C(x)+...\,\,,
$$
where $z,\eta$ are some constants.
The first term is divergent but it is a c-number and it can be ignored if 
we are interested in  operators in normal ordered sense. The second term
is well defined and the others  are equal to zero in the limit 
$x \to x^\prime.$
Our task is to bosonize the single-chain Hamiltonian
$$
H_1=\int dx \left(-iv_F\hat\psi^\dagger\tau_3\partial_1\hat\psi
-\alpha \hat\psi^\dagger\partial_1^2\hat\psi+
g\hat\psi_+^\dagger\hat\psi_+\hat\psi^\dagger_-\hat\psi_-
\right),
 $$
Using the Baker-Hausdorff relation
$$
e^A e^B=e^{A+B} e^{{1\over 2}[A,B]}
$$
we can write the product of two $\psi-$operators as
$$
\hat\psi^\dagger_{\pm}(x)\hat\psi_{\pm}(x_1)=
\pm {1\over{2\pi i(x_1-x)}}: e^{\mp i\left(\hat\Phi_{\pm}(x)-
\hat\Phi_{\pm}(x_1)\right)} : ,
$$
where
$$
\hat\psi_{\pm}(x,t)={1\over{\sqrt{2\pi\delta}}} e^{\pm i\hat\Phi_\pm(x,t)}, $$
$$ \hat \Phi_{\pm}(x,t)=\sqrt\pi\left(\hat\Phi(x,t)
\mp\int_{-\infty}^x\hat \Pi(x^\prime,t)
dx^\prime \right),$$
and $:\,\,\,:$ means normal ordering of operators.
 Here we have rescaled the energy units so that $v_B=1.$ 
Expanding the exponent to the third order in $(x_1-x)$, we get
\begin{eqnarray}
\hat\psi^\dagger_{\pm}(x)\hat\psi_{\pm}(x_1)=
:\pm {1\over{2\pi i}}\biggl[\pm i \hat\Phi_{\pm}^\prime(x)-{1\over 2}
(x_1-x)\Bigl(\bigl(\hat\Phi_\pm^\prime(x)\bigr)^2\mp i
\hat\Phi_\pm^{\prime\prime}(x)\Bigr) \nonumber \\
+{1\over 6}(x_1-x)^2\Bigl(\pm i\hat\Phi_\pm^
{\prime\prime\prime}(x)
-3\Phi_{\pm}^\prime(x)\Phi_{\pm}^{\prime\prime}(x) 
\mp i\bigl(\hat\Phi_\pm^\prime(x)\bigr)^3\Bigr)\biggr]\label{expan}:,
\end{eqnarray}
where all singular c-number terms are ignored and the primes denote 
differentiation with respect to $x.$ Now it is straightforward to
bosonize all operators which we need.
For the terms of the Hamiltonian we have
$$
-iv_F\int dx\hat\psi^\dagger\tau_3\partial_1\hat\psi={1\over 2}\int dx
\Bigl(v_B(\partial_1\Phi)^2+{1\over{v_B}}\Pi^2\Bigr)v_F $$
$$
g \int dx \hat\psi^\dagger_+\hat\psi_+\hat\psi^\dagger_-\hat\psi_-=
{g\over{4\pi}}\int dx \Bigl(v_B(\partial_1\Phi)^2-{1\over{v_B}}\Pi^2\Bigr) 
$$
$$
\int dx \hat\psi^\dagger\partial_1^2\hat\psi=
-{1\over 3}{{\pi}\over{\sqrt{v_B}}}\int dx\,
\partial_1\Phi\Bigl(3\Pi^2+v_B^2(\partial_1\Phi)^2\Bigr), $$
where $v_B$ was restored for a reason which  will be seen below.
 Normal ordering of  operators
is implied hereafter.
In the derivation of the above expressions the terms in (\ref{expan})
 which are total derivatives
were neglected.
The sum of the first and the second terms can be written as the usual 
 free-boson
Hamiltonian if we renormalize the fields
$\Phi\to{\beta\over{\sqrt{\pi}}}\Phi,\Pi\to{{\sqrt{\pi}}\over\beta}\Pi$
so that
$$
-iv_F\int dx\hat\psi^\dagger\tau_3\partial_1\hat\psi+
g \int dx \hat\psi^\dagger_+\hat\psi_+\hat\psi^\dagger_-\hat\psi_-=
\int dx
{1\over 2}
\Bigl(v_B^2(\partial_1\Phi)^2+\Pi^2\Bigr), $$
where 
$$
\beta^4=\pi^2 {{v_F-{g\over{2\pi}}}\over{v_F+{g\over{2\pi}}}},
$$
$$
v_B=\sqrt{ v_F^2-\left({g\over{2\pi}}\right)^2}.
$$
 
Knowing the relation between $v_B$ and $v_F$ we can put $v_B=1$ again.
Note that the above relations depend on a cut-of procedure
and are universal only to the first power in $g.$ 
So the single-chain Hamiltonian  expressed
through the new renormalized fields becomes
\begin{equation}
\hat H_1={1\over 2}\left((\partial_1\hat \Phi)^2
+\hat\Pi^2\right)
+{\alpha\over 3}\beta\partial_1\hat\Phi
\left(3{\pi\over{\beta^2}}\hat\Pi^2+{{\beta^2}\over\pi}
(\partial_1\hat\Phi)^2\right).
\end{equation}
Finally, the current operator expressed through the same fields is
$$
\hat j=-e({{v_F}\over\beta}\hat\Pi+2\alpha \hat \Pi\partial_1\hat\Phi).
$$

 \section*{ Appendix B}

  Lel us prove that to the first order in $\alpha$
 there is no contribution to $\Gamma$ from the first term 
in (\ref{pre})
\begin{equation}
\langle j_0(\xi_3)\psi_{\pm}(\xi_1)\psi_{\pm}^*(\xi_2)\rangle
\langle \psi_{\pm}(\xi_2)\psi_{\pm}^*(\xi_1)\rangle. \label{ac1}
\end{equation}
 To the  zeroth order in $\alpha$ it was already proven  that 
this term gives no contribution to $\Gamma$.
 Therefore we should show that the first order corrections
give no contribution as well. According to (\ref{z}), to the first order in
$\alpha$ we have
\begin{equation}
\langle \psi_{\pm}(\xi_1)\psi_{\pm}^*(\xi_2)\rangle =
G_{\pm}^{(0)}(\xi_1-\xi_2)\left(1+\langle\int d^2\xi
V(\gamma_0+J_0^{\pm},\gamma_1+J_1^\pm)_\xi
\rangle_\gamma \right), \label{ac2}
\end{equation}
where the subscript $\xi$ represents the arguments 
of the $\gamma$-fields
and $J$-functions, for example
$$
V(\gamma_0+J_0^{\pm},\gamma_1+J_1^{\pm})_\xi\equiv
V\Bigl(\gamma_0(\xi)+J_0^\pm(\xi_1,\xi_2,\xi),\gamma_1(\xi)+J_1^\pm
(\xi_1,\xi_2,\xi)\Bigr)
$$
and the symbol $\langle \,\,\,\,\rangle_\gamma $ means  averaging with
the functional 
\begin{equation}
e^{-{1\over 2}\int d^2\xi \gamma^T\hat D^{-1}\gamma}. \label{gf}
\end{equation}
It was also used that $\langle V(\gamma_0,\gamma_1)\rangle_\gamma =0$ to
 get (\ref{ac2}).
 For the correlator containing $j_0$,
using the same formula
to the first order in $\alpha$ we get
\begin{eqnarray}
\langle j_0(\xi_3)\psi_{\pm}(\xi_1)\psi_{\pm}^*(\xi_2)\rangle \nonumber\\ =
-{{e v_F}\over\beta}G_{\pm}^{(0)}(\xi_1-\xi_2)
\left\langle[1+\int d\xi
V(\gamma_0+J_0^{\pm},\gamma_1+J_1^{\pm})_\xi](\gamma_0+J_0^{\pm})_{\xi_3}\right
\rangle_\gamma \nonumber \\
= -{{ev_F}\over\beta} G^{(0)}_{\pm}(\xi_1-\xi_2) 
J_0^{\pm}(\xi_1,\xi_2,\xi_3) \left(1+\langle\int d\xi
V(\gamma_0+J_0^{\pm},\gamma_1+J_1^{\pm})_\xi\rangle_\gamma \right) \nonumber \\
-{{ev_F}\over\beta}
G^{(0)}_{\pm}(\xi_1-\xi_2)\langle \int d\xi
V(\gamma_0+J_0^{\pm},\gamma_1+J_1^{\pm})_\xi
\gamma_0(\xi_3)
\rangle_\gamma ,\label{ac3}
\end{eqnarray}
where the arguments of $J_0,J_1$ are implied according to the same rule as 
above.
Note that the functions $J_0,J_1$ are odd under the exchange of the
arguments $\xi_1$ and $\xi_2$. Therefore the term 
$$
\int d\xi\langle V(\gamma_0+J_0^{\pm},\gamma_1+
J_1^{\pm})_\xi\rangle_\gamma\equiv
\int d\xi\left\langle V\Bigl(\gamma_0(\xi)+J_0^{\pm}(\xi_1,\xi_2,\xi),
\gamma_1(\xi)+
J_1^{\pm}(\xi_1,\xi_2,\xi)\Bigr)\right\rangle_\gamma
$$
is odd under this exchange too.
(One can always change $\gamma_0,\gamma_1\to -\gamma_0,-\gamma_1$ because the 
functional (\ref{gf}) is invariant under this transformation.)
 One can see that after the substitution of 
(\ref{ac2},\ref{ac3}) into (\ref{ac1}) the correction 
in (\ref{ac2}) and the correction of the same kind in  (\ref{ac3}) cancel 
each other
(to the first order) and (\ref{ac1}) becomes
\begin{eqnarray}
\langle j_0(\xi_3)\psi_{\pm}(\xi_1)\psi_{\pm}^*(\xi_2)\rangle_\gamma
\langle \psi_{\pm}(\xi_2)\psi_{\pm}^*(\xi_1)\rangle_\gamma 
= -{{ev_F}\over\beta} G^{(0)}_{\pm}(\xi_1-\xi_2) G^{(0)}_{\pm}(\xi_2-\xi_1)
J_0^{\pm}(\xi_1,\xi_2,\xi_3) \nonumber \\
 -{{ev_F}\over \beta}G^{0}_{\pm}(\xi_1-\xi_2)G^{0}_{\pm}(\xi_2-\xi_1)
\langle \int d\xi
V(\gamma_0+J_0^{\pm},\gamma_1+J_1^{\pm})_\xi\gamma_0(\xi_3)\rangle_\gamma.
\label{ac4}
\end{eqnarray}
The first term in this expression is the zero order term which gives
no contribution to $\Gamma.$ The second term gives zero
when integrated over $x_3$.
To see this we need the correlation functions of the $\gamma$-fields.
Taking the Fourier transformation of the Green function (\ref{gd}) one  gets
$$
\langle\gamma_1(\xi_1)\gamma_1(\xi_2)\rangle_\gamma=
\langle\gamma_0(\xi_1)\gamma_0(\xi_2)\rangle_\gamma=
{1\over{4\pi}}\left({1\over{(t+ix)^2}}+{1\over{(t-ix)^2}}\right),
$$
$$
\langle\gamma_1(\xi_1)\gamma_0(\xi_2)\rangle_\gamma=
\langle\gamma_0(\xi_1)\gamma_1(\xi_2)\rangle_\gamma=
-{1\over{4\pi}}\left( {1\over{(t+ix)^2}}-{1\over{(t-ix)^2}}\right).
$$
From the form of the $\gamma$-correlators
written above one can see that
$$
\int dx_1\langle \gamma(\xi_1)\gamma^\prime(\xi_2)\rangle=0,
$$
where $\gamma$ and $\gamma^\prime$ are any fields from the  $\gamma_1,
\gamma_2$-fields.
Therefore there is no contribution  to $\Gamma$ from the term (\ref{ac1}).


\begin{thebibliography}{99}
\bibitem{shankar}
R. Shankar, {\it Low-Dimensional Quantum Field Theories for
Condensed Matter Physicists} ( World Scientific Publishing 
Co. Pte. Ltd., 1992), p. 353
\bibitem{fradkin}
E. Fradkin, {\it Field Theories of Condensed Matter Systems} 
(Addison-Wesley,Redwood City, 1991)
\bibitem{wen}
X.G. Wen, Phys. Rev. B {\bf 42}, 6623 (1990).
\bibitem{haldane}
F.D.M. Haldane, J. Phys. C {\bf14}, 2585 (1981).
\bibitem{schulz}
H.J. Schulz, Int.J.Mod.Phys. B {\bf5}, 57 (1991).
\bibitem{jerome}
D. Jerome and H.J.Schulz, Adv. Phys. {\bf31}, 299 (1982).
 \bibitem{peskin}
M.E. Peskin and D.V. Schroeder, {\it An Introduction to Quantum Field Theory}
 (Addison-Wesley Publishing Company, 1995).
\bibitem{braz}
S. Brazovskii and V.Yakovenko, Sov. Phys. JETP {\bf 62}, 1340 (1985)
\bibitem{bourb}
D.Boies, C.Bourbonnais and A.-M.S.Tremblay, Phys. Rev. Lett. {\bf 74}, 
968 (1995)
\bibitem{yakov}
V.M. Yakovenko, JETP Lett. {\bf 56}, 510 (1992)
\bibitem{gor}
L.P. Gor'kov, Sov. Phys. Usp. {\bf 27}, 809 (1984)   
\end{thebibliography}
 \end{document}